\documentclass{appolb}%
\usepackage{epsfig}
\usepackage{rotating}
\usepackage{hyperref}
\usepackage{comment,color}
\usepackage{braket}
\usepackage{amssymb,amsmath,bm}
\usepackage{setspace,lipsum}
\usepackage{amsmath}
\usepackage{amsfonts}
\usepackage{amssymb}
\usepackage{graphicx}%
\providecommand{\U}[1]{\protect\rule{.1in}{.1in}}
\begin{document}

\title{Companion poles: from the $a_{0}(980)$ to the $X(3872)$}
\author{F.~Giacosa$^{a,b}$
\address{$^a$Institute of Physics, Jan Kochanowski University, 25-406 Kielce, Poland}
\address{$^b$Institut f\"ur Theoretische Physik, Johann Wolfgang Goethe-
Universit\"at, 60438 Frankfurt am Main, Germany} }
\maketitle

\begin{abstract}
When an unstable ordinary quark-antiquark state couples strongly to other
low-mass mesons (such as pions, kaons, $D$-mesons, etc.), the quantum
fluctuations generated by the decay products dress the bare `seed' $\bar{q}q$
state and modify its spectral functions. The state is associated to a pole on
the complex plane. When the coupling to the decay products is sufficiently
large, a remarkable and interesting phenomenon takes place: dynamically
generated companion states (or poles) might emerge. Some resonances listed in
the PDG , such as the $a_{0}(980),$ the $K_{0}^{\ast}(700),$ and the
$X(3872),$ can be well understood by this mechanism, that we briefly review in
these proceedings. On the other hand, we show that the $Y(4008)$ and $Y(4260)$
are not independent resonances (or poles), but manifestations of $\psi(4040)$
and $\psi(4160),$ respectively.

\end{abstract}


\section{Introduction}

The idea behind companion poles is quite simple: in the easiest scenario, one
starts with a \textit{single} bare field with quantum numbers $J^{PC}$ which
corresponds to a well-defined $\bar{q}q$ state with a certain bare mass,
typically very close to the predictions of the quark model \cite{isgur}. Then,
a Lagrangian in which this \textquotedblleft seed\textquotedblright\ state
couples strongly to some standard mesons (such a pions, kaons, $D$-mesons,
$\rho$-mesons, ...) is written down. As a consequence, mesonic loops dress the
original state. The original pole on the real axis moves down in the complex
plane. Moreover, when the interaction is strong enough, other poles can
appear: these are dynamically generated companion poles. In some cases, such
poles can be interpreted as additional resonances and some of the
supernumerary states listed in the PDG \cite{pdg} can have such an origin. A
general feature of companion poles is that they fade away in the large-$N_{c}%
$limit \cite{lebed}: in fact, the coupling of the original $\bar{q}q$ state to
other ordinary mesons scales as $1/\sqrt{N_{c}},$ therefore the quantum
fluctuations become smaller for and additional poles do not emerge.

The mechanism outlined above was described in\ Refs. \cite{boglione,tornqvist}
in the context of light scalar mesons. Later on, the concept of companion
poles has been revisited in detail in Ref. \cite{a0}, in which the
$a_{0}(980)$ is described as a companion pole of a predominantly standard
$\bar{q}q$ resonance $a_{0}(1450)$ and where a detailed comparison with the
previous works of Ref. \cite{boglione,tornqvist} is presented. In the last
four years, the approach has been studied for various states
\cite{soltysiak,3770,4040,4160,x3872}, as we shall discuss in more detail in
the next section (see Table 1). For related ideas about the emergence of
companion poles, we refer also to \cite{dullemond,coito3872,xiao,pelaezrev}
and refs. therein.

\section{Companion poles: status}

In this section, we describe the present status of the approach and the
results obtained through its applications to resonances both in the light and
in the charmonium sectors.

In order to explain the idea, we consider two explicit examples. In the first
one, the seed state is the scalar state $K_{0}^{\ast}$, which decays to $K\pi$
and -after dressing- mainly corresponds to $K_{0}^{\ast}(1430)$; in the second
case, the seed state is the $\bar{c}c$ bare field $\psi_{\mu},$ which decays
into $DD$ and predominantly corresponds to $\psi(3770)$. The Lagrangians for
these two systems are:
\begin{align}
\mathcal{L}_{K_{0}^{\ast}}  &  =aK_{0}^{\ast-}\pi^{0}K^{+}+bK_{0}^{\ast
-}\partial_{\mu}\pi^{0}\partial^{\mu}K^{+}+...\text{ ,}\label{lagk}\\
\mathcal{L}_{\psi}  &  =ig_{\psi}\psi_{\mu}\left(  \partial^{\mu}D^{+}%
D^{-}-\partial^{\mu}D^{-}D^{+}\right)  +...\text{ ,} \label{lagpsi}%
\end{align}
where the dots refer to other isospin combinations, see details in
\cite{soltysiak,3770}. In both cases, the decay widths as function of the
`running' mass of the decaying particle can be expressed as:
\begin{equation}
\Gamma_{_{K_{0}^{\ast}}}(m)=\Gamma_{_{K_{0}^{\ast}}}^{\text{tl}}(m)F_{\Lambda
}(m)\text{ ; }\Gamma_{_{\psi}}(m)=\Gamma_{_{\psi}}^{\text{tl}}(m)F_{\Lambda
}(m)\text{ ,}%
\end{equation}
where the tree-level part (tl) is obtained from the standard Feynman rules for
the local Lagrangian in Eqs. (\ref{lagk})-(\ref{lagpsi}), while the quantity
$F_{\Lambda}(m)$ is a vertex function which takes into account the finite
dimensions of the mesons. It could be formally introduced already at the
Lagrangian level by rendering it nonlocal \cite{nonlocal}. The function
$F_{\Lambda}(m)$ should guarantee convergence of the loops, thus $F_{\Lambda
}(m\rightarrow\infty)=0$ sufficiently fast. A typical choice, valid in the
reference frame of the decaying particle, is $F_{\Lambda}(m)=e^{-2k(m)/\Lambda
},$ where $k(m)$ is the modulus of the three-momentum of one of the outgoing
decay products and $\Lambda\simeq0.5$ GeV is the typical energy scale for the
overlap of extended mesons. Note, even if the vertex function cuts the
three-momentum $k$, Lorentz invariance is guaranteed \cite{covariance}.

One consider the propagator of the seed state dressed by loops of the decay
products. Its scalar part is
\begin{equation}
\Delta_{j}^{-1}(m)=m^{2}-M_{0,j}^{2}+\Pi_{j}(m^{2})\text{ , }j=K_{0}^{\ast
},\psi\text{ , }%
\end{equation}
where $\Pi_{j}(m^{2})$ is the loop function such that $\operatorname{Im}%
\Pi_{j}(m^{2})=m\Gamma_{j}(m)$. Since the imaginary part is known, the loop
function $\Pi_{j}(m^{2})$ can be obtained by dispersion relations. In the
first Riemann sheet (IRS), $\Pi_{j}(m^{2})$ is regular everywhere, a part from
a cut along the real axis. When the coupling constant(s) is (are) sent to
zero, the so-called seed pole of $\Delta_{j}(m)$ is $m_{seed}=M_{0}%
-i\varepsilon.$ For nonzero couplings, the seed pole $m_{seed}$ moves down in
the IIRS: the pole mass is \ $\operatorname{Re}[m_{seed}]$ (usually, not far
from $M_{0}$) and the decay width is $-2\operatorname{Im}[m_{seed}]$. But for
coupling constant large enough, there can be a second, dynamically generated
companion pole:%
\begin{equation}
m_{comp}\text{ such that }\Delta_{j}^{-1}(m_{comp})_{IIRS}=0\text{ and
}m_{comp}\neq m_{seed}\text{ .}%
\end{equation}
The companion poles has a completely different `movement' on the complex
plane: for small coupling, it lies very far from the real axis and then it
approaches the real axis from below when the coupling increases. Eventually,
for very large coupling, it can be even closer to the real axis than the
original seed pole. (This is not the case for the two examples above, but it
applies for the $a_{0}$-system, see below).

Next, one defines the spectral function as \cite{salam,lupo}
\begin{equation}
d_{j}(m)=\frac{2m}{\pi}\operatorname{Im}[\Delta_{j}(m)]\text{ }\rightarrow
\int_{0}^{\infty}d_{j}(m)dm=1\text{ .}%
\end{equation}
The normalization is a crucial property, since it allows to interpret
$d_{j}(m)$ as a mass probability density (for a detailed proof, see Refs.
\cite{lupoprd}). Note, even when the companion pole is present, strictly
speaking there is only one `state' properly normalized to unity. Typically,
the companion pole generates an enhancement of the spectral function at low
energies (or even a second peak as for $a_{0}(980)$ and $X(3872)$). Note, the
here outlined approach is valid at the (resummed) one-loop level. Fortunately,
it seems to be a good approximation in hadron physics \cite{schneitzer}.

In Table 1 we report the present status of some resonances: for given quantum
numbers, the bare fields with the spectroscopic notation and $\bar{q}q$
content, the resulting predominantly $q\bar{q}$ resonances, and the companion
poles are listed. Then, below the Table we briefly discuss each case separately.

\newpage

\begin{center}
\textbf{Table 1}: Summary of light and heavy systems in which companion poles
have been investigated.%

\begin{tabular}
[c]{|c|c|c|c|c|c|}\hline
$J^{PC}$ & $%
\begin{array}
[c]{c}%
\text{Bare field}\\
n^{2S+1}L_{J}\\
\bar{q}q
\end{array}
$ & $%
\begin{array}
[c]{c}%
\text{Main}\\
\text{decays}%
\end{array}
$ & $%
\begin{array}
[c]{c}%
\text{Predom. }\bar{q}q\\
\text{pole (GeV)}%
\end{array}
$ & $%
\begin{array}
[c]{c}%
\text{Companion}\\
\text{pole (GeV)}%
\end{array}
$ & Ref.\\\hline
$0^{++}$ & $%
\begin{array}
[c]{c}%
a_{0}\\
1^{3}P_{0}\\
u\bar{d},..
\end{array}
$ & $%
\begin{array}
[c]{c}%
KK\\
\pi\eta,\pi\eta^{\prime}%
\end{array}
$ & $%
\begin{array}
[c]{c}%
a_{0}(1450)\\%
\begin{array}
[c]{c}%
1.456\\
-i0.134
\end{array}
\end{array}
$ & $%
\begin{array}
[c]{c}%
a_{0}(980)\\%
\begin{array}
[c]{c}%
0.970\\
-i0.045
\end{array}
\end{array}
$ & \cite{a0}\\\hline
$0^{++}$ & $%
\begin{array}
[c]{c}%
K_{0}^{\ast}\\
1^{3}P_{0}\\
u\bar{s},..
\end{array}
$ & $K\pi$ & $%
\begin{array}
[c]{c}%
K_{0}^{\ast}(1430)\\%
\begin{array}
[c]{c}%
1.413\\
-i0.127
\end{array}
\end{array}
$ & $%
\begin{array}
[c]{c}%
K_{0}^{\ast}(700)\\%
\begin{array}
[c]{c}%
0.746\\
-i0.262
\end{array}
\end{array}
$ & \cite{soltysiak}\\\hline
$1^{--}$ & $%
\begin{array}
[c]{c}%
\psi\\
1^{3}D_{1}\\
c\bar{c}%
\end{array}
$ & $DD$ & $%
\begin{array}
[c]{c}%
\psi(3770)\\%
\begin{array}
[c]{c}%
3.777\\
-i0.0123
\end{array}
\end{array}
$ & $%
\begin{array}
[c]{c}%
-\\%
\begin{array}
[c]{c}%
3.741\\
-0.0018
\end{array}
\end{array}
$ & \cite{3770}\\\hline
$1^{--}$ & $%
\begin{array}
[c]{c}%
\psi\\
3^{3}S_{1}\\
c\bar{c}%
\end{array}
$ & $%
\begin{array}
[c]{c}%
DD,DD^{\ast}\\
D^{\ast}D^{\ast},D_{s}D_{s}\\
D_{s}^{\ast}D_{s}%
\end{array}
$ & $%
\begin{array}
[c]{c}%
\psi(4040)\\%
\begin{array}
[c]{c}%
4.053\\
-i0.039
\end{array}
\end{array}
$ & $%
\begin{array}
[c]{c}%
-\\%
\begin{array}
[c]{c}%
3.934\\
-i0.030
\end{array}
\end{array}
$ & \cite{4040}\\\hline
$1^{--}$ & $%
\begin{array}
[c]{c}%
\psi\\
2^{3}D_{1}\\
c\bar{c}%
\end{array}
$ & $%
\begin{array}
[c]{c}%
DD,DD^{\ast}\\
D^{\ast}D^{\ast},D_{s}D_{s}\\
D_{s}^{\ast}D_{s},D_{s}^{\ast}D_{s}^{\ast}%
\end{array}
$ & $%
\begin{array}
[c]{c}%
\psi(4160)\\%
\begin{array}
[c]{c}%
4.199\\
-i0.033
\end{array}
\end{array}
$ & $%
\begin{array}
[c]{c}%
-\\
-
\end{array}
$ & \cite{4160}\\\hline
$1^{++}$ & $%
\begin{array}
[c]{c}%
\chi_{c,1}(2P)\\
2^{3}P_{1}\\
c\bar{c}%
\end{array}
$ & $DD^{\ast}$ & $%
\begin{array}
[c]{c}%
\chi_{c,1}(2P)(?)\\%
\begin{array}
[c]{c}%
3.995\\
-i0.036
\end{array}
\end{array}
$ & $%
\begin{array}
[c]{c}%
X(3872)\\%
\begin{array}
[c]{c}%
3.87164\\
-i\varepsilon\text{ {\small (virtual)}}%
\end{array}
\end{array}
$ & \cite{x3872}\\\hline
\end{tabular}

\bigskip
\end{center}

\begin{itemize}
\item $a_{0}(980)$ and $a_{0}(1450)$ \cite{a0}. One starts with a unique field
$a_{0}$ with a bare mass of about 1.2 GeV coupled to light mesons according to
the constrains of chiral symmetry \cite{dick}. Then, upon including the loops,
$a_{0}(1450)$ is predominantly $\bar{q}q$, and the resonance $a_{0}(980)$ is a
dynamically generated companion pole (for a detailed discussion of light
scalar mesons, see \cite{pelaezrev}).\ Quite remarkably, the loops are so
strong that the corresponding spectral function $d_{a_{0}}(m)$ contains two peaks.

\item $K_{0}^{\ast}(700)$ and $K_{0}^{\ast}(1430)$ \cite{soltysiak}: the seed
state $K_{0}^{\ast}$ lies well above 1 GeV. $K_{0}^{\ast}(1430)$ corresponds
to the (dressed) $\bar{q}q$ state, while $K_{0}^{\ast}(700)$ is dynamically
generated. In the spectral function there is no peak for this state, but a
slight low-energy enhancement. Recently, the existence of this
non-conventional meson has been in the centre of many investigations, e.g.
Ref. \cite{pelaezk}. The PDG2018 hag re-named this state as $K_{0}^{\ast
}(700)$ (previously, $K_{0}^{\ast}(800)$) and included in the summary table.
Our study clearly confirms the existence of this state and provides a clear
physical interpretation of its nature.

\item $\psi(3770)$ \cite{3770}: the non-Breit-Wigner form of the spectral
function is caused by the loops. Also in this case two poles appear. Yet, the
dynamically generate pole is quite close to the seed one, hence no new name
for an independent state is assigned.\ 

\item $\psi(4040)$ \cite{4040}: the bare $\bar{c}c$ state couples strongly to
various $D$ mesons. The spectral function is strongly distorted and two poles
are generated, just as for $\psi(3770).$ The dynamically generated pole does
\textbf{not} correspond to the enhancement $Y(4008)$ \cite{belle}. A broad
distorted resonance-like structure may emerge in the $j/\psi\pi\pi$ channel
through the process $\psi(4040)\rightarrow DD^{\ast}\rightarrow j/\psi\pi\pi$
(because the real part of the loop $DD^{\ast}$ is peaked at the \ $DD^{\ast}$
threshold). Hence, $Y(4008)$ is not an independent state, but a $DD^{\ast}%
$-loop manifestation of $\psi(4040)$.

\item $\psi(4160)$ \cite{4160} (actual mass: $4.191$ GeV \cite{pdg}): in this
case there is a unique (relevant) pole. As before, the chain $\psi
(4160)\rightarrow D_{s}^{\ast}D_{s}^{\ast}\rightarrow j/\psi\pi\pi$ generates
a resonance-like structure peaked at about $4.222$ GeV (this is the
$D_{s}^{\ast}D_{s}^{\ast}$ threshold where again the real part of the loop is
enhanced). This signal can be assigned to the $Y(4260)$ (also called
$\psi(4260)$ in the PDG \cite{pdg}). Then, $Y(4260)$ is not an independent
resonance, but a loop manifestation of the state $\psi(4160)$ (and of its
pole) shifted of about $40$ MeV in mass.

\item $X(3872)$ and $\chi_{c1}(2P)$ \cite{x3872}: a bare seed state $\chi
_{c1}(2P)$ gets dressed by $DD^{\ast}$ loops. At the lowest $D_{0}D_{0}^{\ast
}$ threshold the spectral function develops a very high and narrow peak: the
$X(3872)$. In the complex plane, there is a virtual pole just below
$D_{0}D_{0}^{\ast}$. The state $\chi_{c1}(2P)$ has a well-defined pole, but
the corresponding peak can fade away, explaining the difficulty to measure it
in experiments.
\end{itemize}

\section{Conclusions}

In this work we have briefly reviewed the concept dynamical generation of
companion poles and the status of some resonances on the basis of this idea.
Other resonances could also emerge as companion poles, as for instance the
state $D_{s}(2317)$. Moreover, as for the $Y(4008)$ and $Y(4260)$, other
enigmatic $Y$ states (see \cite{rev2016} for a review) could be not real, but
`loop' manifestation of conventional $\bar{c}c$ states.

\bigskip

\textbf{Acknowledgments}: The author thanks S. Coito, M. Piotrowska, T.
Wolkanowksi-Gans, and D. H.\ Rischke for cooperations leading to the
publications listed in Table 1. Moreover, the author acknowledges support from
the Polish National Science Centre NCN through the OPUS projects no.
2015/17/B/ST2/01625 and no. 2018/29/B/ST2/02576.

\end{document}